# A Brewster route to Cherenkov detectors

Xiao Lin[1,2,#], Hao Hu[3,#], Sajan Easo[4], Yi Yang[5], Yichen Shen[6], Kezhen Yin[7], Michele Piero Blago[8,9], Ido Kaminer[10,*], Baile Zhang[11,12,*], Hongsheng Chen[1,2], John Joannopoulos[5], Marin Soljačić[5], and Yu Luo[3,*]

[1]*Interdisciplinary Center for Quantum Information, State Key Laboratory of Modern Optical Instrumentation, ZJU-Hangzhou Global Science and Technology Innovation Center, College of Information Science and Electronic Engineering, Zhejiang University, Hangzhou 310027, China.*
[2]*International Joint Innovation Center, ZJU-UIUC Institute, Zhejiang University, Haining 314400, China.*
[3]*School of Electrical and Electronic Engineering, Nanyang Technological University, Nanyang Avenue, Singapore 639798, Singapore.*
[4]*Particle Physics Department, Rutherford-Appleton Laboratory (STFC), Didcot, OX110QX, UK.*
[5]*Research Laboratory of Electronics and Department of Physics, Massachusetts Institute of Technology, Cambridge, Massachusetts 02139, USA.*
[6]*Lightelligence, 268 Summer Street, Boston, Massachusetts 02210, USA.*
[7]*Mantaline Corporation, 4754 E High Street, Mantua, OH44106, USA.*
[8]*European Organization for Nuclear Research (CERN), Geneva 1211, Switzerland.*
[9]*Cavendish Laboratory, University of Cambridge, Cambridge CB3 0HE, UK.*
[10]*Department of Electrical Engineering, Technion-Israel Institute of Technology, Haifa 32000, Israel.*
[11]*Division of Physics and Applied Physics, School of Physical and Mathematical Sciences, Nanyang Technological University, 21 Nanyang Link, Singapore 637371, Singapore.*
[12]*Centre for Disruptive Photonic Technologies, Nanyang Technological University, Singapore 637371, Singapore.*
[#]*These authors contributed equally.*
[*]E-mail: kaminer@technion.ac.il (I. Kaminer); blzhang@ntu.edu.sg (B. Zhang); luoyu@ntu.edu.sg (Y. Luo).

**The Cherenkov effect enables a valuable tool, known as the Cherenkov detector, to identify high-energy particles via the measurement of the Cherenkov cone. However, the sensitivity and momentum coverage of such detectors are intrinsically limited by the refractive index of the host material. Especially, identifying particles with energy above multiple gigaelectronvolts requires host materials with a near-unity refractive index, which are often limited to large and bulky gas chambers. Overcoming this fundamental material limit is important for future particle detectors yet remains a long-standing scientific challenge. Here, we propose a different paradigm for Cherenkov detectors that utilizes the broadband angular filter made from stacks of variable one-dimensional photonic crystals. Owing to the Brewster effect, the angular filter is transparent only to Cherenkov photons from a precise incident angle, and particle identification is achieved by mapping each Cherenkov angle to the peak-intensity position of transmitted photons in the detection plane. This unique property of the angular filter is exceptionally beneficial to Cherenkov detection as it enables the realization of a non-dispersive pseudo refractive index over the entire visible spectrum. Moreover, such a pseudo refractive index can be flexibly tuned to arbitrary values,**

**including those close to unity. Our angular-selective Brewster paradigm offers a feasible solution to implement compact and highly sensitive Cherenkov detectors especially in beam lines and it can cover a wide momentum range using readily available dielectric materials.**

A charged particle travelling in a transparent host medium emits photons when it travels faster than the phase velocity of the photons in that medium. This phenomenon is known as Cherenkov radiation, which is first observed experimentally by P. A. Cherenkov (under the guidance of S. Vavilov) [1,2] and later interpreted theoretically by I. M. Frank and I. Tamm [3,4]. Remarkably, Cherenkov radiation [5-8] has enabled the invention of Cherenkov detectors [9-14] for identifying particles over a large momentum range in high-energy physics and astrophysics. The Cherenkov detector has played an essential role during the discovery of many elementary particles, including anti-protons [15], J/$\psi$ particles [16], neutrino oscillations [17], etc.

According to Frank and Tamm's theory of Cherenkov radiation [18], the particle velocity $v$ can be determined by measuring the light emission angle $\theta_{CR}$ (also known as the Cherenkov angle)

$$\cos\theta_{CR} = \frac{c}{nv}, \tag{1}$$

where $n$ is the refractive index of the host medium, and $c$ is the speed of light in vacuum. Although larger refractive indices give rise to lower Cherenkov thresholds and higher photon yield, they are not always desirable in Cherenkov detectors. The reason is that large refractive index decreases the sensitivity of the Cherenkov angle to small changes in the velocity. In general, the highest detection sensitivity is obtained at $\theta_{CR} \to 0$ or $n \to c/v$ (The detection sensitivity is defined as $\frac{d\theta_{CR}}{dv} = \frac{n}{c} \cdot \frac{\cos^2\theta_{CR}}{\sin\theta_{CR}}$, which reaches its maximum at $\theta_{CR} \to 0$). In other words, to identify high-energy particles (i.e. $v \to c$), transparent dielectrics with a near-unity refractive index are often required. Such a constraint limits the host materials used in many Cherenkov radiators to low-index. This is for example the case in ring imaging Cherenkov (RICH) detectors that use gases for the identification of particles with momenta larger than 10 GeV/$c$ [14,19,20].

Another limitation in state-of-the-art Cherenkov detectors is related to the photon emission efficiency: the use of low-index materials inevitably leads to low efficiency (i.e. the photon yield is proportional to $1 - \frac{c^2}{n^2 v^2}$ according to Frank and Tamm's theory). Especially when operating near-threshold (i.e. $v \to c/n$), the photon yield approaches zero. Consequently, traditional gas radiators generally require bulky gas chambers to produce sufficient photons for detection [14]. Regular dielectrics offer a possible route to increase the photon emission efficiency, but their large refractive indices (generally far above unity) effectuate different types of high-energy particles to generate the same Cherenkov angle, namely $\theta_{CR} \to \cos^{-1}(1/n)$, which thus makes particle identification impossible. As an example, quartz has a refractive index around 1.4, and its corresponding momentum coverage is limited below 3.5 GeV/$c$ [21]. All high-energy particles with momenta above this value emit light at $\theta_{CR} \in [44.30°, 44.42°]$, independent of the particle velocity and particle type. Under this scenario, measuring $\theta_{CR}$ cannot lead to the identification of the corresponding particle type.

Recently, several theoretical attempts have been made to relax the material limitations in Cherenkov detectors by using modern concepts from nanophotonics and metamaterials [22,23]. One attempt proposes metal-based anisotropic metamaterials with one component of the effective refractive index close to unity [22]. Another study makes use of all-dielectric one-dimensional (1D) photonic crystals, where the constructive interference of resonance transition radiation is adopted to control the effective Cherenkov angle [23]. These nanophotonic Cherenkov detectors can achieve an enhanced sensitivity for any desired momenta range, however only at a specific working frequency. In fact, the working frequency is the major drawback of all nanophotonics-based Cherenkov detectors, i.e. they all have a narrow working bandwidth, resulting from the inherent chromatic dispersion of the constitutive materials (e.g., metal) and the resonant nature of periodic structures (e.g., photonic crystals). There are many more recent advances and ongoing efforts on the Cherenkov effect in nanophotonic settings and in novel material platforms [24-35]. Nevertheless, the design of a *broadband* Cherenkov detector using regular transparent dielectrics and at the same time with enhanced performance has remained a long-standing scientific challenge.

Here we propose a new paradigm for Cherenkov detectors by exploiting a broadband angular filter. This broadband angular filter is comprised of stacks of many 1D photonic crystals of different periodicities but identical constituent materials [Fig. S1a]. As a result, the Brewster effect makes the angular filter transparent only to *p*-polarized (i.e. transverse magnetic, TM) light incident at the Brewster angle, while the light with other polarizations or incident at other angles is totally reflected. Moreover, we can readily tailor the band gaps of these 1D photonic crystals to cover a broad spectral range, thus also making the angular selectivity broadband, spanning the entire visible spectrum [36-39]. After transmitting through the broadband angular filter, the Cherenkov radiation in the detection plane features a pseudo Brewster-Cherenkov angle $\theta_{\text{BCR}}$, namely the angle between the particle velocity and the tangential wavevector parallel to the detection plane [Fig. 1]. Remarkably, we find that the measurement of $\theta_{\text{BCR}}$ can provide an approach for particle identification at any momentum coverage with wide bandwidth and high sensitivity. This approach thus can tackle the key drawback listed above for nanophotonic Cherenkov detectors. Our approach is especially useful for the identification of a beam of charged particles with different momenta, even when the flux of charged particles is high.

We now proceed to analyze the essential role of the Brewster effect in our proposed particle detectors. As schematically shown in Fig. 1, we consider the charged particle travelling at a constant velocity $v$ in a host material with the relative permittivity of $\varepsilon_{\text{h}}$ and along a trajectory parallel to the surface of the broadband angular filter. Without loss of generality, we set the broadband angular filter composed of two regular transparent dielectrics with relative permittivities $\varepsilon_{\text{r1}}$ and $\varepsilon_{\text{r2}}$ [Fig. S1a]. The detection plane is located at $y = y_0$ beneath the broadband angular filter, and it is parallel to but far away from the particle trajectory [Fig. 1]. The Cherenkov radiation transmitting through the broadband angular filter has a tangential wavevector $\bar{k}_{\text{BCR}} = \hat{x}k_x + \hat{z}k_z$ parallel to the detection plane. On the one hand, $k_z = \omega/v$ is fixed by the kinematic feature of the charged particle [23,40]. On the other hand, the magnitude of $\bar{k}_{\text{BCR}}$ is locked by the intrinsic electromagnetic property of the broadband angular filter. To be specific, only the *p*-polarized light incident at the Brewster angle can transmit through our angular filter for the entire spectral range of 400 to 700 nm [Figs. S1-S3], and our particle detector will operate over this broadband wavelength range.

According to the Brewster effect, $|\bar{k}_{\text{BCR}}| = \sqrt{\frac{\varepsilon_{r1}\varepsilon_{r2}}{\varepsilon_{r1}+\varepsilon_{r2}}}\frac{\omega}{c}$ [18], which enables us to define a *pseudo refractive index*

$$n_{\text{BCR}} = \sqrt{\frac{\varepsilon_{r1}\varepsilon_{r2}}{\varepsilon_{r1}+\varepsilon_{r2}}}. \qquad (2)$$

We highlight that the broadband angular filter is treated as a realistic multilayered structure, instead of an effectively homogeneous material, and hence $n_{\text{BCR}}$ [see Methods for its derivations] is completely different from the effective refractive index obtained using the standard homogenization theory. Interestingly, $n_{\text{BCR}}$ is related to the pseudo Brewster-Cherenkov angle $\theta_{\text{BCR}}$ [see inset of Fig. 1] by the following equation,

$$\cos\theta_{\text{BCR}} = \frac{k_z}{|\bar{k}_{\text{BCR}}|} = \frac{c}{n_{\text{BCR}}v}. \qquad (3)$$

Equation (3) generalizes the regular Frank-Tamm formula (i.e. equation (1)), providing a general route to engineer the Cherenkov radiation through the Brewster effect. In what follows, we shall explore the potential applications of this generalized Frank-Tamm formula for particle identification. As our approach exploits the Brewster effect, we refer to the particle detector depicted in Fig. 1 as the Brewster-Cherenkov detector.

To facilitate the conceptual demonstration, Fig. 2 plots the spatial distribution of Cherenkov radiation in the detection plane. After penetrating the broadband angular filter, the transmitted Cherenkov photons feature two tails, which are symmetric with respect to the *z*-axis [Figs. 2a-d & S4-S5; movie S1]. The pseudo Brewster-Cherenkov angle, and hence the particle velocity, can be further determined with high sensitivity by measuring the peak-intensity position of Cherenkov radiation in the detection plane [Figs. 2e & S5]. To illustrate the distinct advantages of our Brewster-Cherenkov detector, we also plot the Cherenkov radiation in the detection plane without the angular filter [Fig. 2(f-i)]. Through the comparison, we find that the pseudo Brewster-Cherenkov angle $\theta_{BCR}$ [Fig. 2a-d] is much more sensitive to the variation in particle velocity than the regular Cherenkov angle $\theta_{CR}$ [Fig. 2f-i]. This remarkable property clearly demonstrates that the broadband angular filter can effectively improve the sensitivity of Cherenkov detectors.

We highlight that such sensitivity improvement occurs over a broadband frequency range because the pseudo refractive index $n_{BCR}$ in the generalized Frank-Tamm formula can be made approximately non-dispersive using regular transparent dielectrics that have a negligible dispersion in the visible range [41]. To illustrate this point, we plot $n_{BCR}$ as a function of the Brewster angle. As shown in Fig. 3, $n_{BCR}$ can be flexibly engineered to arbitrary values (including those close to unity) by a suitable choice of $\varepsilon_{r1}$ and $\varepsilon_{r2}$ for the two constituent materials of the broadband angular filter. For example, according to equation (2), $\varepsilon_{r1} = 2.18$ (e.g., SiO$_2$) and $\varepsilon_{r2} = 3.07$ (Al$_2$O$_3$) [41] give rise to a pseudo refractive index $n_{BCR} = 1.13$, which is much smaller (and hence more close to unity) than the lowest refractive index found in natural solid materials (i.e. 1.37 for MgF$_2$) [19,20,41-44]. A pseudo refractive index even closer to unity can be achieved using other material combinations. For example, $n_{BCR} = 1.0026$, if taking polymers with $\varepsilon_{r1} = 1.8578$ (tetrafluoroethylene-co-hexafluoropropylene-co-vinylidene fluoride (THV, 3M Dyneon 221AZ)) and $\varepsilon_{r2} = 2.1904$ (poly(methyl methacrylate), namely PMMA) [45-46].

A key feature of Brewster-Cherenkov detectors is that the pseudo refractive index $n_{BCR}$ determines their sensitivity and momentum range. Figure 4a shows the relation between the pseudo Brewster-Cherenkov angle $\theta_{BCR}$ and particle momenta for four elementary particles, namely electron, pion, kaon, and proton. In this exemplary case, we take $n_{BCR} = 1.13$ and a particle momentum fixed at 2 GeV/$c$. The resulting values of $\theta_{BCR}$ are 27.8°for electron, 27.5° for pion, 24.3° for kaon, and 12.2° for proton [Fig. 4a]. Such a variation in $\theta_{BCR}$ indicates that $n_{BCR} = 1.13$ is suitable for the identification of particles with a momentum less than 10 GeV/c. In comparison, $n_{BCR} = 1.0026$ gives rise to a Brewster-Cherenkov detector capable of identifying particles with a momentum larger than 10 GeV/$c$. More interestingly, when $n_{BCR} = 1.000001$, the corresponding Brewster-Cherenkov detector can even identify particles with ultra-high momenta in the TeV/$c$ range. These results clearly demonstrate that the proposed Brewster-Cherenkov detector can achieve arbitrary momentum coverage with high sensitivity through a proper design of the pseudo refractive index.

The pseudo Brewster-Cherenkov angle and the particle velocity can be determined by directly measuring the peak-intensity position $x_{BCR}$ of Cherenkov radiation in the detection plane [Figs. 4b & S6-S7]. Mathematically, we have $\left|\frac{x_{BCR}}{y_0}\right| = \frac{n_{BCR} \cdot \sin\theta_{BCR}}{\sqrt{\varepsilon_{r1} - n_{BCR}^2}} + \Delta(y_0)$, where $\Delta(y_0)$ is the normalized displacement resulting from the refraction of light through the broadband angular filter. For simplicity but without loss of generality, we choose $n_{BCR} = 1.13$ and $y_0 = 2.3$ mm, and plot in Fig. 4b the ratio $\left|\frac{x_{BCR}}{y_0}\right|$ as a function of momentum for the four elementary particles. At the fixed momentum of 2 GeV/$c$, the ratio $\left|\frac{x_{BCR}}{y_0}\right|$ gradually varies from 0.49, 0.48, 0.43 to 0.22 for electrons, pions, kaons and protons, respectively [e.g., see the corresponding intensity distributions of Cherenkov radiation in the detection plane in Fig. 2a-e].

As a final remark, we highlight that our Brewster approach has several unique advantages over the traditional methods for particle identification. First, the Brewster approach eliminates the strict requirement of near-unity-index host materials for the design of Cherenkov radiators. In other words, distinct from traditional Cherenkov detectors such as the RICH detector [13-17], our Brewster approach does not have any special requirements on the refractive index of the host material where the particle is travelling, since the sensitivity of Brewster-Cherenkov detectors is directly determined by $n_{BCR}$ of the broadband angular filter. Consequently, high-index transparent solids or gases with low atomic numbers can now be used as the host material. Such a high-index host material can significantly enhance the number of photons penetrating the broadband angular filter and reaching the detection plane, enabling a higher-efficiency Cherenkov detector.

Another important advantage of our Brewster approach in comparison with all previous nanophotonic approaches [22,23] is that the broadband angular filter does not need to be placed in the path of the high-energy particle beam. The charged particles can travel at a large distance away from the surface of the broadband angular filter so that all the Cherenkov photons are produced in the host material in which the particle is travelling. This way, the generation of secondary particles from the broadband angular filter can be effectively reduced. Last but not least, the performance of the proposed particle detector is robust to fabrication imperfections and geometric fluctuations of the broadband angular filter (see analysis in Refs. [36-39]). Therefore, this Brewster approach provides promising options to facilitate the design of advanced Cherenkov detectors with enhanced sensitivity, large bandwidth, miniaturized size, ultralight weight and wide momentum coverage, all using readily available regular dielectrics. These enhanced capabilities are especially attractive in the identification of high-energy particles in beam lines.

## Acknowledgements
The work at Zhejiang University was sponsored by the National Natural Science Foundation of China (NNSFC) under Grants No. 61625502, No.11961141010, and No. 61975176, the Top-Notch Young Talents Program of China, and the Fundamental Research Funds for the Central Universities. Y.L. was sponsored in part by Singapore Ministry of Education (No. MOE2018-T2-2-189 (S), MOE2017-T1-001-239 (RG91/17 (S)), A*Star AME Programmatic Funds (No. A18A7b0058) and National Research Foundation Singapore Competitive Research Program (No. NRF-CRP18-2017-02). B.Z. was sponsored in part by Singapore Ministry of Education (No. MOE2018-T2-1-022 (S), MOE2016-T3-1-006 and Tier 1 RG174/16 (S)). I.K. was sponsored in part by the Azrieli Faculty Fellowship, the Israel Science Foundation (Grant No. 831/19), and the European Research Council (Starter Grant no. 851780). This material is based upon work supported in part by the U.S. Army Research Office through the Institute for Soldier Nanotechnologies, under Contract No. W911NF-18-2-0048, and by the Binational USA-Israel Science Foundation (BSF) 2018288.

## Author contributions
All authors contributed extensively to this work. X.L. initiated the idea. H.H. and X.L. performed the calculation. S.E., Y.Y., Y.S., K.Y., M.P.B., H.C., I.K., B.Z., J.J., M.S., and Y.L. analyzed data and interpreted detailed results. X.L., H.H., Y.L., S.E., I.K., and B.Z. wrote the manuscript with input from the others. Y.L., I.K., S.E., B.Z., H.C., J.J., and M.S. supervised the project.

## Methods
**Calculation of light emission from a charged particle moving parallel to the surface of the broadband angular filter.** The basic structural setup is shown in Fig. 1. All the fields of emitted light are calculated in the framework of electromagnetic wave theory by applying the method of plane wave expansion. We let the *z*-axis parallel to the moving direction of charged particle in Fig. 1; as such, the induced current density is $\bar{J}(\bar{r},t) = \hat{z}qv\delta(x)\delta(y)\delta(z-vt)$. The field distribution in each region is obtained through matching the boundary conditions or by using the transfer matrix method. The detailed procedure of calculation is shown in section S1 of the supporting information.

**Derivation of the pseudo refractive index in equation (2) of the main text.** For conceptual demonstration, here we set region 1 having the relative permittivity of $\varepsilon_{r1}$ and region 2 having the relative permittivity of $\varepsilon_{r2}$. By enforcing the electromagnetic boundary condition at the interface between regions 1 & 2, we can readily obtain the Brewster angle $\theta_{\text{Brewster}}$, at which the reflection for *p*-polarized waves is zero [18]. That is, according to the Brewster effect, the Brewster angle for *p*-polarized waves in region 1 has $\tan\theta_{\text{Brewster}} = \sqrt{\varepsilon_{r2}/\varepsilon_{r1}}$ [18]. At this Brewster angle, the wavevector component of light parallel to the interface has $k_{\text{BCR}} = k_1 \sin\theta_{\text{Brewster}}$, where $k_1 = \sqrt{\varepsilon_{r1}}\omega/c$ is the wavevector of light in region 1 and $\sin\theta_{\text{Brewster}} = \sqrt{\varepsilon_{r2}}/\sqrt{\varepsilon_{r1}+\varepsilon_{r2}}$. In other words, we have $k_{\text{BCR}} = \frac{\omega}{c}\sqrt{\varepsilon_{r1}} \cdot \frac{\sqrt{\varepsilon_{r2}}}{\sqrt{\varepsilon_{r1}+\varepsilon_{r2}}} = \frac{\omega}{c}\sqrt{\frac{\varepsilon_{r1}\varepsilon_{r2}}{\varepsilon_{r1}+\varepsilon_{r2}}}$. Since we denote $k_{\text{BCR}} = n_{\text{BCR}}\frac{\omega}{c}$, we directly have equation (2) in the main text, namely $n_{\text{BCR}} = \sqrt{\frac{\varepsilon_{r1}\varepsilon_{r2}}{\varepsilon_{r1}+\varepsilon_{r2}}}$. Due to the momentum matching at each interface, the value of $k_{\text{BCR}}$ is the same for different regions in the broadband angular filter, when the light transmits through the angular filter. In addition, we highlight that all the calculations in this work treat the broadband angular filter as a realistic layered structure, instead of an effectively homogenized material by using the effective medium theory. These discussions are shown in section S2 of the supporting information.

**Design of the broadband angular filter.** The broadband angular filter [Fig. S1a] is comprised of many stacks (i.e. $M$ stacks) of 1D photonic crystals. All 1D photonic crystals are made of two regular transparent dielectrics, which have a relative permittivity of $\varepsilon_{r1}$ and $\varepsilon_{r2}$, respectively. The $i^{\text{th}}$ 1D photonic crystal has a pitch of $P_i = d_{i1} + d_{i2}$, where $d_{i1}$ and $d_{i2}$ are the thickness of two dielectric slabs in each pitch. All these 1D photonic crystals have a pitch number of $N$ and a thickness ratio of $d_{i1}/d_{i2} = 3/2$. This way, the band gap of each 1D photonic crystal can be flexibly tunable by changing $P_i$. The light transmission through the broadband angular filter is both angle-dependent and polarization-dependent. For the $p$-polarized light with arbitrary incident angle (except the one equal to the Brewster angle), the light is almost fully reflected by judiciously overlapping the band gaps of these 1D photonic crystals [Figs. S1 & S2]. For the $p$-polarized light incident at the Brewster angle, it can safely pass through the broadband angular filter with no reflection [Figs. S1 & S2]. For the $s$-polarized light, the transmission through the designed broadband angular filter is negligible for arbitrary incident angle [Fig. S3]. The detailed design strategy is shown in sections S2-S4 of the supporting information.

**Cherenkov radiation in the detection plane.** For Cherenkov radiation passing through the broadband angular filter, while the direction of their in-plane wavevector $\bar{k}_{\text{BCR}} = \hat{x}k_x + \hat{z}k_z$ has a pseudo Brewster-Cherenkov angle with respect to the particle trajectory [Fig. S4], their motion in the detection plane is parallel to the particle trajectory (i.e. along the $z$ axis); see the discussion of Cherenkov radiation in the detection plane in section S5 of the supporting information and their dynamics in the detection plane in Movie S1.

**Peak-intensity position of Cherenkov radiation in the detection plane at $y = y_0$.** The normalized peak-intensity position $\left|\frac{x_{\text{BCR}}}{y_0}\right| = \frac{n_{\text{BCR}} \cdot \sin\theta_{\text{BCR}}}{\sqrt{\varepsilon_{r1} - n_{\text{BCR}}^2}} + \Delta(y_0)$ is analytically calculated according to the ray tracing theory. Here, $\Delta(y_0) = -\left(\frac{\sqrt{n_{\text{BCR}}^2 - \frac{c^2}{v^2}}}{\sqrt{\varepsilon_{r1} - n_{\text{BCR}}^2}} - \frac{\sqrt{n_{\text{BCR}}^2 - \frac{c^2}{v^2}}}{\sqrt{\varepsilon_{r2} - n_{\text{BCR}}^2}}\right)\frac{y_b}{y_0}$ is the normalized displacement induced by the refraction of light through the broadband angular filter, where $y_b$ is the total thickness of dielectric regions with $\varepsilon_{r2}$ in the designed broadband angular filter. Since $\Delta(y_0) \propto \frac{y_b}{y_0}$, we have $\Delta(y_0) \to 0$ if $y_b \ll y_0$. In other words, if the detection plane is far away from the particle trajectory and if their distance is much larger than the finite thickness of the broadband angular filter, we have $\left|\frac{x_{\text{BCR}}}{y_0}\right| = \frac{n_{\text{BCR}} \cdot \sin\theta_{\text{BCR}}}{\sqrt{\varepsilon_{r1} - n_{\text{BCR}}^2}}$. The detailed discussion of $\left|\frac{x_{\text{BCR}}}{y_0}\right|$ for a fixed broadband angular filter under different values of $y_0$ is shown in section S6 and Figs. S5-S7 in the supporting information.

**Robustness of the performance of Brewster-Cherenkov detectors to particle's trajectory.** The Brewster-Cherenkov detector has the potential to infer the projection of particle trajectory in the $xz$ plane (or the detection plane at $y = y_0$). This is because the intensity distribution of the transmitted Cherenkov radiation in the detection plane is symmetric with respect to the projection of the particle trajectory in the $xz$ plane [Fig. 2a-e]. Due to this unique feature, the sensitivity of Brewster-Cherenkov detectors is in principle insensitive to the direction of particle velocity, if the particle velocity is parallel to the surface of the broadband angular filter. On the other hand, for the Brewster-Cherenkov detector, we can always set the particle trajectory very far away from the top surface of the broadband angular filter. Then if the particle velocity has a very small angle with respect to the surface of the broadband angular filter (but the particle would not penetrate through the filter), the performance of Brewster-Cherenkov detectors would not be degraded, since the feature of the transmitted Cherenkov radiation in the detection plane is mostly preserved.

**Influence of the finite thickness of the broadband angular filter on the performance of Brewster-Cherenkov detectors.** More discussions on the performance of Brewster-Cherenkov detectors are provided in section S7 and Figs. S8-S9 of the supporting information. When the stack number $M$ of 1D photonic crystals and the periodicity number $N$ of each 1D photonic crystal are finite, the *p*-polarized light incident at the angles very close to the Brewster angle can also safely pass through the broadband angular filter. This way, there is a small angular (and thus spatial) spread of the transmitted Cherenkov radiation in the detection plane, such as those shown in Fig. S1b-e. This phenomenon would degrade the sensitivity of Brewster-Cherenkov detectors. However, the sensitivity of Brewster-Cherenkov detector can still be guaranteed by effectively avoiding this phenomenon, through increasing both the values of $M$ and $N$ in the practical implementation [Fig. S9].

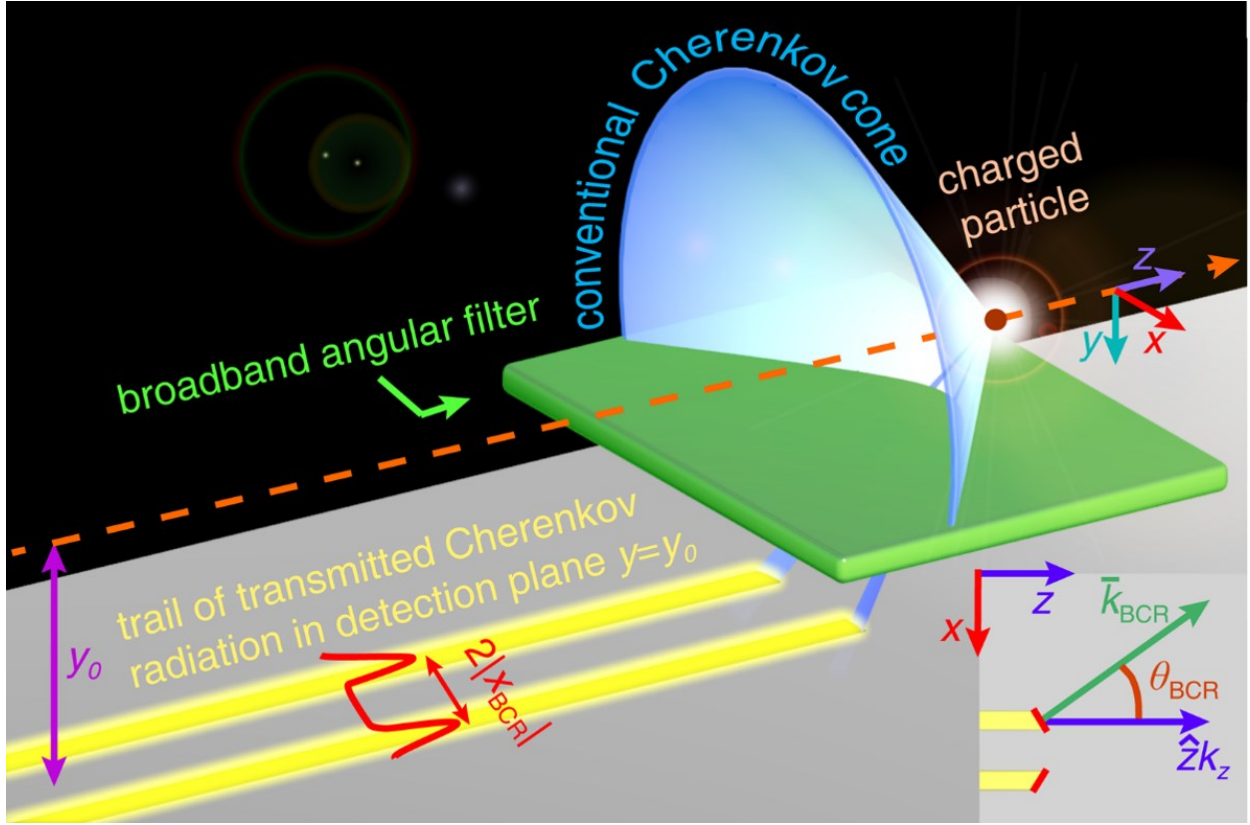

**Figure 1 | Schematic of the proposed Brewster-Cherenkov detector based on a broadband angular filter.** In our design, the charged particle moves along a trajectory far away from the surface of the broadband angular filter. The broadband angular filter is designed by exploiting the Brewster effect in stacks of many 1D photonic crystals with various periodicities but two same constituent materials. After the angular filtering, the transmitted Cherenkov radiation in the detection plane has $\cos\theta_{\text{BCR}} = k_z/|\bar{k}_{\text{BCR}}|$ [inset], where $k_z = \omega/v$ and $\bar{k}_{\text{BCR}}$ is the tangential wavevector of light parallel to the detection plane. The measurement of the peak-intensity position $|x_{\text{BCR}}/y_0|$ of Cherenkov radiation in the detection plane provides the information of the pseudo Brewster-Cherenkov angle $\theta_{\text{BCR}}$ and the particle velocity, and it thus can be exploited for particle identification.

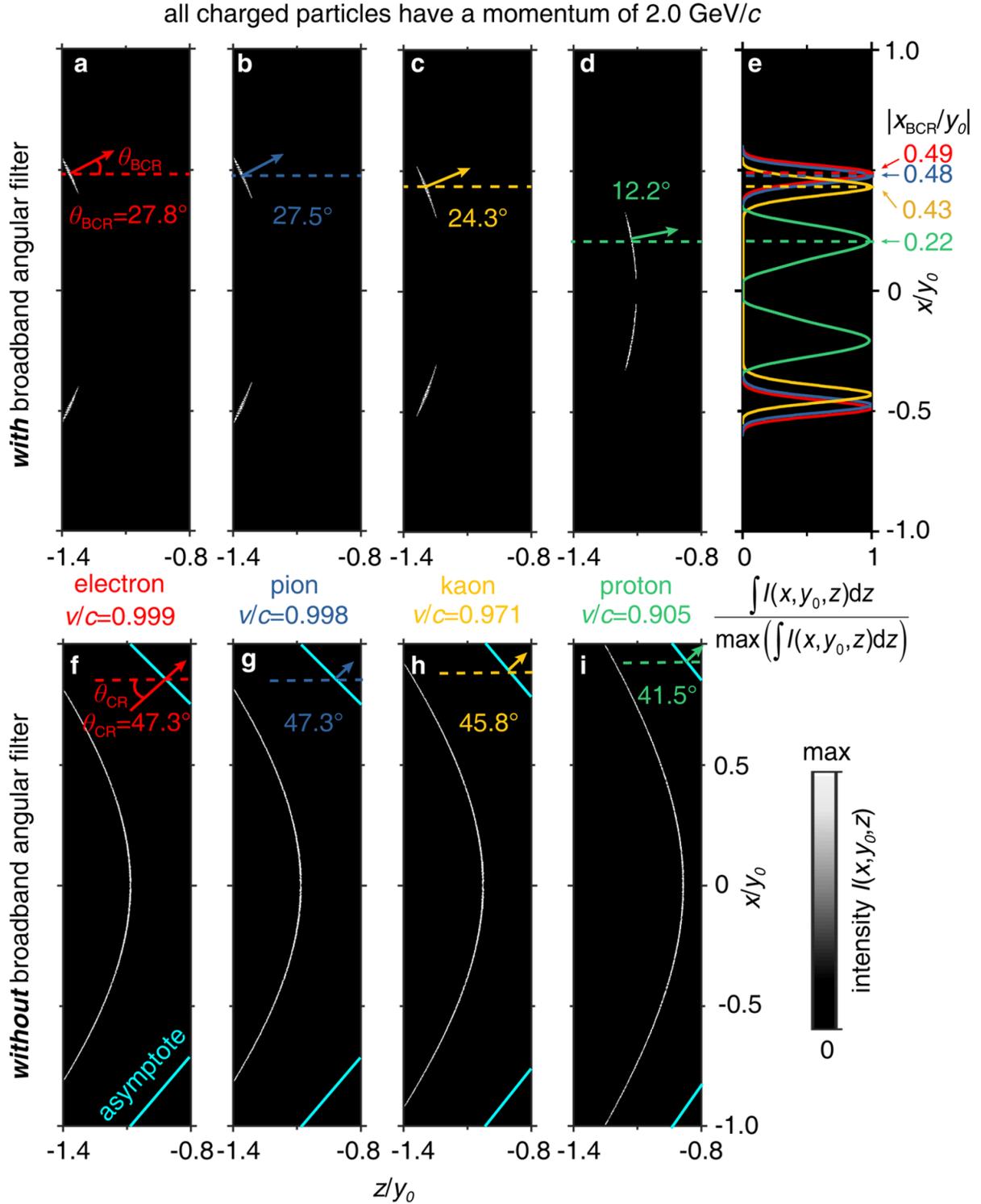

**Figure 2 | Cherenkov radiation in the detection plane of Brewster-Cherenkov detectors.** Here we plot the intensity distribution of Cherenkov radiation for four elementary charged particles with a fixed momentum, when these charged particles arrive at $z = 0$. **a-e**, The Brewster-Cherenkov detector has the

broadband angular filter [Fig. 1]. **f-i**, For comparison, the angular filter is removed. The Cherenkov radiation in (a-d) is related to the pseudo Brewster-Cherenkov angle $\theta_{\text{BCR}}$, whose value can be obtained through the measurement of the peak-intensity position in (e), while the asymptotic line of Cherenkov radiation in (f-i) is related to the regular Cherenkov angle $\theta_{\text{CR}}$. Our proposed approach has no critical requirement on the permittivity of the host material in which the charged particle is travelling (e.g., $\varepsilon_{\text{h}} = \varepsilon_{\text{r1}}$ used in the calculation). Here and below, the considered wavelength varies from 400 nm to 700 nm. In addition, $y_0 = 2.3$ mm, $\varepsilon_{\text{r1}} = 2.18$ and $n_{\text{BCR}} = 1.13$.

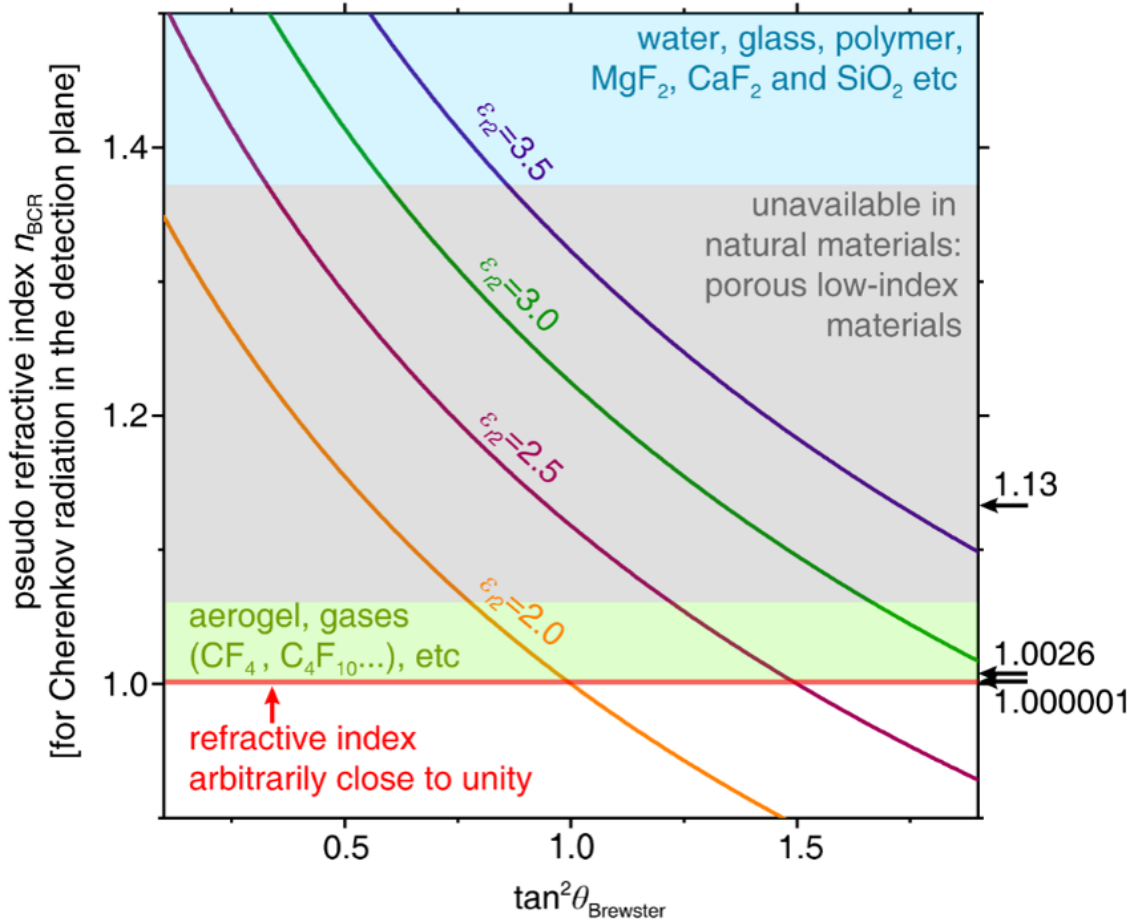

**Figure 3 | Engineering the pseudo refractive index $n_{BCR}$ for Brewster-Cerenkov detectors, achieving parameter regimes that do not exist in natural materials.** For the transmitted Cherenkov radiation in the detection plane, its wavevector component parallel to the detection plane has $k_{BCR} = n_{BCR}\omega/c$, where $n_{BCR} = \sqrt{\frac{\varepsilon_{r1}\varepsilon_{r2}}{\varepsilon_{r1}+\varepsilon_{r2}}}$. Here we plot $n_{BCR}$ as a function of the Brewster angle $\theta_{Brewster}$ under different values of $\varepsilon_{r2}$, where $\tan\theta_{Brewster} = \sqrt{\varepsilon_{r2}/\varepsilon_{r1}}$.

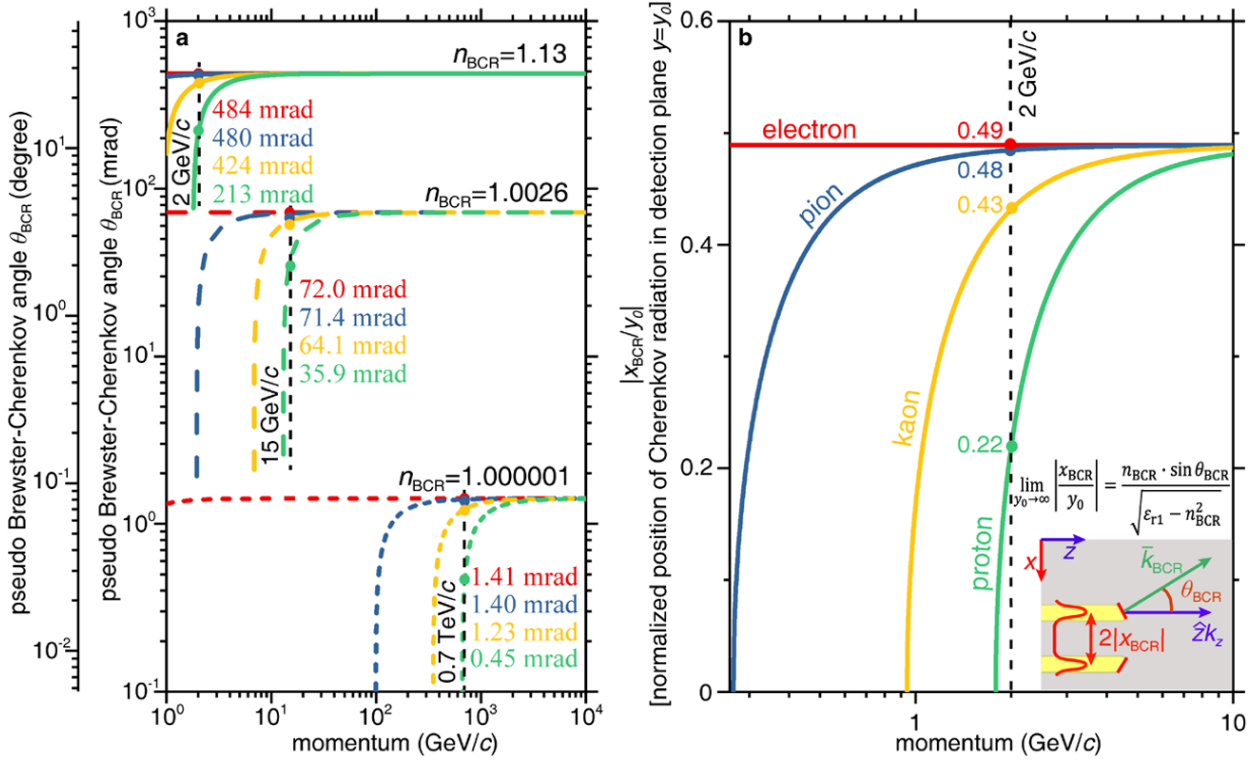

**Figure 4 | Performance of Brewster-Cherenkov detectors in the identification of high-energy particles. a**, Pseudo Brewster-Cherenkov angle $\theta_{\text{BCR}}$ versus particle momentum. This figure is plotted according to the generalized Frank-Tamm formula $\cos\theta_{\text{BCR}} = \frac{c}{n_{\text{BCR}}v}$, by transforming the particle velocity to the momentum. The values of pseudo Brewster-Cherenkov angles for four elementary particles, namely electron (red), pion (blue), kaon (yellow) and proton (green), with a momentum of 2 GeV/$c$, 15 GeV/$c$ or 700 GeV/$c$ are given in the plot. **b**, Normalized peak-intensity position $|x_{\text{BCR}}/y_0|$ of the transmitted Cherenkov radiation in the detection plane versus particle momentum. The measurement of $|x_{\text{BCR}}/y_0|$ provides the information of $\theta_{\text{BCR}}$ and further the particle velocity. In (b), $y_0 = 2.3$ mm and $n_{\text{BCR}} = 1.13$.